\documentclass[]{pasj01}

\begin{document} 
\Received{}
\Accepted{}

\title{
A search for optical transients associated 
with Fast Radio Burst 150418\thanks{
Based on data collected at Subaru Telescope, 
which is operated by the National Astronomical Observatory of Japan.}\thanks{
Based on observations obtained at the Gemini Observatory 
acquired through the Gemini Observatory Archive,  
which is operated by the Association of Universities for Research in Astronomy, Inc., 
under a cooperative agreement with the NSF on behalf of the Gemini partnership: 
the National Science Foundation (United States), 
the National Research Council (Canada), CONICYT (Chile), 
Ministerio de Ciencia, Tecnología e Innovación Productiva (Argentina), 
and Ministério da Ciência, Tecnologia e Inovação (Brazil).} 
}

\author{
Yuu \textsc{Niino}\altaffilmark{1}, 
Nozomu \textsc{Tominaga}\altaffilmark{2,3}, 
Tomonori \textsc{Totani}\altaffilmark{4,5}, 
Tomoki \textsc{Morokuma}\altaffilmark{6}, 
Evan \textsc{Keane}\altaffilmark{7}, 
Andrea \textsc{Possenti}\altaffilmark{8}, 
Hajime \textsc{Sugai}\altaffilmark{3}, 
Shotaro \textsc{Yamasaki}\altaffilmark{4}
}
\altaffiltext{1}{National Astronomical Observatory of Japan, 
2-21-1 Osawa, Mitaka, Tokyo 181-8588, Japan}
\altaffiltext{2}{Department of Physics, Faculty of Science and Engineering, 
Konan University, 8-9-1 Okamoto, Kobe, Hyogo 658-8501, Japan}
\altaffiltext{3}{Kavli Institute for the Physics and Mathematics of the Universe (WPI), 
The University of Tokyo Institutes for Advanced Study, 5-1-5 Kashiwa, Chiba 277-8583, Japan}
\altaffiltext{4}{Department of Astronomy, School of Science, The University of Tokyo, 
7-3-1 Hongo, Bunkyo-ku, Tokyo 113-0033, Japan}
\altaffiltext{5}{Research Center for the Early Universe, School of Science, 
The University of Tokyo, 7-3-1 Hongo, Bunkyo-ku, Tokyo 113-0033, Japan}
\altaffiltext{6}{Institute of Astronomy, Graduate School of Science, The University of Tokyo, 
2-21-1 Osawa, Mitaka, Tokyo 181-8588, Japan} 
\altaffiltext{7}{SKA Organisation, Jodrell Bank Observatory, Cheshire SK11 9DL, UK}
\altaffiltext{8}{INAF-Osservatorio Astronomico di Cagliari, 
Via della Scienza 5, I-09047 Selargius (CA), Italy}
\email{yuu.niino@nao.ac.jp}


\maketitle

\begin{abstract}
We have searched for optical variability in 
the host galaxy of the radio variable source which is possibly 
associated with fast radio burst (FRB) 150418. 
We compare images of the galaxy taken 1 day after the burst using Subaru/Suprime-Cam 
with images taken $\sim$ 1 year after the burst using Gemini-South/GMOS. 
No optical variability is found 
between the two epochs with a limiting 
absolute magnitude $\gtrsim -19$ (AB). 
This limit applies to optical variability of the putative active galactic nucleus 
in the galaxy on a timescale of $\sim$ 1 year, 
and also to the luminosity of an optical counterpart 
of FRB~150418 one day after the burst should it have occurred in this galaxy. 
\end{abstract}

\section{Introduction}

Fast radio burst (FRB) 150418 was detected 
by the Parkes radio telescope at 04:29:07 on 18 April 2015 (UTC, \cite{Keane:2016a}). 
A multiwavelength follow up campaign was conducted 
with various telescopes including the Australia Telescope 
Compact Array (ATCA, 5.5 GHz and 7.5 GHz) and Subaru (optical, $r$- and $i$-band). 
A fading radio object with a negative spectral index ($f_\nu\propto\nu^{-1.37}$) 
was detected by ATCA within the error circle of FRB~150418 in the first 6 days after the burst. 
This lead to a claimed association between the source and FRB~150418, 
however, it is possible that the fading source is scintillation of radio emission 
from an active galactic nucleus (AGN) and unrelated with FRB~150418 
\citep{Williams:2016a, Akiyama:2016a, Johnston:2017a}.

Optical imaging observations of the error circle of FRB~150418 
using Suprime-Cam \citep{Miyazaki:2002a} on the Subaru telescope 
were conducted 1 to 2 days after the burst. 
Although no peculiar variable object was found within the error circle, 
an early type galaxy was clearly detected 
at the position of the fading object observed by ATCA. 
The galaxy is also detected by the WISE satellite \citep{Wright:2010a} 
and catalogued as WISE J071634.59--190039.2 
(hereafter WISE~J0716--19). 
The subsequent spectroscopy of WISE~J0716--19 using Subaru/FOCAS \citep{Kashikawa:2002a} 
revealed that its redshift is $z = 0.492\pm0.008$ \citep{Keane:2016a}. 

No variable object was found in WISE~J0716--19 
in the optical images taken with Suprime-Cam 1 to 2 days after the burst. 
However, an optical counterpart of FRB~150418 
might be missed by those observations even if it existed at the time of observation, 
if the variability timescale of the optical counterpart is longer than the observation period. 
In this study, we compare the images taken 1 to 2 days after the burst 
with images of the same field taken $\sim$ 1 year after the burst 
using GMOS on Gemini-South \citep{Hook:2004a}, 
to search for any optical 
transient event that may have occurred in WISE~J0716--19 
during the period between the two observations. 
Throughout the paper, we assume the fiducial cosmology 
with $\Omega_{\Lambda}=0.7$, $\Omega_{m}=0.3$, 
and $H_0=$ 70 km s$^{-1}$ Mpc$^{-1}$. 
Magnitudes are given in the AB system. 

\section{Data}

Our optical follow up observations of FRB~150418 using Subaru/Suprime-Cam 
were performed on 19 and 20 April 2015 (UTC, \cite{Keane:2016a}, 
hereafter the event images). 
To detect any optical variability of WISE~J0716--19 
with longer timescale than $\sim$ 1 day, 
we retrieved GMOS observations of WISE~J0716--19 conducted $\sim$ 1 year after the burst 
from the Gemini observatory archive as reference images
(Program ID: GS-2016A-Q-104). 
The reference images were taken under lightly cloudy conditions 
(CC = 70\%-tile\footnote{
http://www.gemini.edu/sciops/telescopes-and-sites/observing-condition-constraints}). 

The event images are reduced using 
the Hyper-Suprime-Cam pipeline version 3.8.5 \citep{Bosch:2018a}, 
which is based on the LSST pipeline \citep{Ivezic:2008a, Axelrod:2010a}, 
and the reference images are reduced using 
PyRAF/IRAF\footnote{PyRAF is a product of the Space Telescope 
Science Institute, which is operated by AURA for NASA. 
IRAF is distributed by the National Optical Astronomy Observatories, 
which are operated by the Association 
of Universities for Research in Astronomy, Inc., 
under cooperative agreement with the National Science Foundation.}, 
together with the Gemini IRAF package. 

\begin{figure*}
 \begin{center}
  \includegraphics[width=15cm]{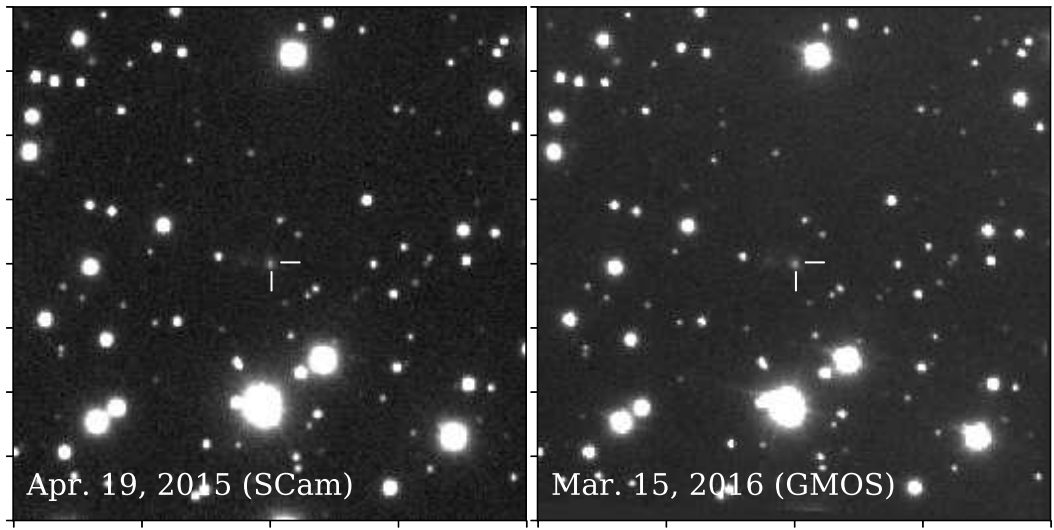}
 \end{center}
 \caption{
   {\it Left panel}: the \timeform{80''} $\times$ \timeform{80''} field image in $i$-band 
    centered on WISE~J0716--19 which is highlighted with cross hairs. North is up, East to the left. 
    The image is taken on 19 Apr. 2015 using Subaru/Suprime-Cam (the event image). 
   {\it Right panel}: same as the left panel but taken on 15 Mar. 2016 using Gemini-South/GMOS (the reference image). 
    The pixels are aligned with those of the event image using the \texttt{remap} program in WCSTools. 
 }\label{fig:field}
\end{figure*}

We summarize information of the observations in Table~\ref{tab:obs}. 
In the following discussions, 
we use the event images obtained on 19 April 2015 
and the reference images obtained on 15 March 2016, 
due to the poor seeing conditions on 20 April 2015 and 11 April 2016. 
The 80$\times$80 arcsec$^2$ field centered on WISE~J0716--19 
in $i$-band is shown in Figure~\ref{fig:field}. 
We calibrate the flux scale of the event images using unsaturated objects 
in the same field that are catalogued in the Pan-STARRS1 
database \citep{Chambers:2016a} as photometric standards. 

\begin{table*}
  \tbl{Observations of WISE~J0716--19.}{%
  \begin{tabular}{lllll}
  \hline\noalign{\vskip3pt} 
    Start time (UTC) & Telescope/instrument & Filter & Exposures & Seeing \\
  \hline\noalign{\vskip1pt} 
  \hline\noalign{\vskip2pt} 
    19 Apr. 2015 05:58:27 & Subaru/Suprime-Cam & $i$-band & 60 sec $\times$ 10 & \timeform{0''.7} \\ 
    19 Apr. 2015 06:25:07 & Subaru/Suprime-Cam & $r$-band & 60 sec $\times$ 15 & \timeform{0''.7} \\
    20 Apr. 2015 05:35:39 & Subaru/Suprime-Cam & $i$-band & 60 sec $\times$ 20 & \timeform{0''.9} \\
    20 Apr. 2015 06:15:46 & Subaru/Suprime-Cam & $r$-band & 60 sec $\times$ 20 & \timeform{1''.2} \\
    15 Mar. 2016 02:13:08 & Gemini-South/GMOS & $r$-band & 150 sec $\times$ 7 & \timeform{0''.7} \\
    15 Mar. 2016 02:35:14 & Gemini-South/GMOS & $i$-band & 150 sec $\times$ 7 & \timeform{0''.6} \\
    11 Apr. 2016 00:08:20 & Gemini-South/GMOS & $z$-band & 150 sec $\times$ 7 & \timeform{0''.9} \\
    11 Apr. 2016 00:30:34 & Gemini-South/GMOS & $i$-band & 150 sec $\times$ 7 & \timeform{0''.9} \\
  \hline\noalign{\vskip3pt} 
  \end{tabular}}\label{tab:obs}
\end{table*}

\section{Search for a variable object}

\subsection{Relative photometry between the two epochs}
\label{ssec:photo}

To achieve accurate relative photometry between the two epochs, 
we compare photon counts of unsaturated objects in the field, 
and calibrate the flux scale of the reference images 
so that the fluxes of the unsaturated objects 
are the same as those in the event images. 
We perform photometry of objects in the images 
using the SExtractor software \citep{Bertin:1996a}. 

\begin{figure*}
 \begin{center}
  \includegraphics[width=15cm]{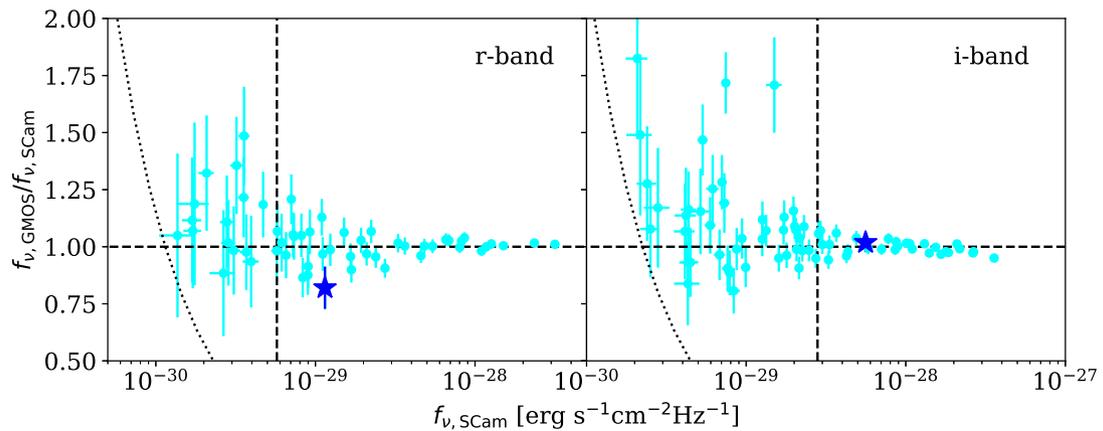}
 \end{center}
 \caption{
    The flux ratios of objects in the vicinity of WISE~J0716--19 between the event 
    and reference images in $r$- and $i$-band (the left and right panels, respectively). 
    The error bars are $1\sigma$ significance. 
    WISE~J0716--19 is shown with a star symbol. 
    The vertical dashed line indicates the lower flux limit
    above which objects are used for the calibration of relative photometry, 
    and the horizontal dashed line indicates $f_{\nu,{\rm GMOS}}/f_{\nu,{\rm SCam}}=1.0$. 
    The dotted curve represents a constant $f_{\nu,{\rm GMOS}}$. 
    One of the two outliers with $f_{\nu,{\rm GMOS}}/f_{\nu,{\rm SCam}}>1.5$ 
    at $f_{\nu,{\rm SCam}} \sim 10^{-29}$ erg s$^{-1}$cm$^{-2}$Hz$^{-1}$ 
    in the right panel is a diffuse object which may suffer 
    from uncertainty in the aperture determination, 
    and the other one is blended with a nearby bright object. 
 }\label{fig:flux}
\end{figure*}

In Figure~\ref{fig:flux}, we show the flux ratios of 
the unsaturated objects between the event and reference images 
as a function of their flux densities in the event image.
As expected, the flux ratio of fainter objects are more scattered. 
Furthermore, there is a systematic error where faint objects 
appear systematically brighter in the reference image in $i$-band. 
To avoid any unwanted impact of faint objects on the photometry, 
we use objects at least 50\% as bright as 
WISE~J0716--19 for the photometric calibration. 

WISE~J0716--19 is shown with a star symbol in Figure~\ref{fig:flux}. 
Although significant change in the flux density 
of WISE~J0716--19 is not found in $i$-band, 
the flux density has decreased in $r$-band by 20\% between the two epochs. 
The measured flux densities of WISE~J0716--19 
in the event and reference images by SExtractor 
are $(1.15\pm0.07)\times10^{-29}$ and $(0.94\pm0.09)\times10^{-29}$ 
erg s$^{-1}$cm$^{-2}$Hz$^{-1}$, respectively. 
However, this difference likely results 
from extra errors in the photometry
that are not taken into account in the error estimation by SExtractor, 
such as uncertainty of aperture determination. 
We have executed SExtractor independently on the event and reference images 
because the pixel alignments are different between the images, 
and the elliptical aperture for WISE~J0716--19 
determined by SExtractor is different in each image. 
To examine the dependence of the flux density 
on the determination of the photometric aperture, 
we perform photometry of WISE~J0716--19 in $r$-band 
with circular apertures of various diameters 
between \timeform{3''} and \timeform{7''} with a sampling rate of \timeform{0''.1}, 
instead of the elliptical aperture determined by SExtractor. 

The mean and the root-mean-square error of the flux densities 
obtained in this range of aperture diameters are $(1.13\pm0.13)\times10^{-29}$ 
erg s$^{-1}$cm$^{-2}$Hz$^{-1}$ in both of the event and reference images. 
Thus, we conclude that the decrease of the flux density 
in $r$-band seen in Figure~\ref{fig:flux} is not real. 
We also note that WISE~J0716--19 is an extended source 
while most of other objects in the field are point sources, 
and hence it suffers more from the uncertainty 
of the aperture determination than other objects, 
and a faint object that resides $\sim$ \timeform{5''} east 
of WISE~J0716--19 may also affect the photometry. 

\subsection{Image subtraction}

To search for a transient object in WISE~J0716--19, 
we subtract the calibrated reference images from the event images. 
We use the \texttt{remap} program in WCSTools\footnote{
http://tdc-www.cfa.harvard.edu/software/wcstools/} 
to align the pixels of the reference images obtained 
using GMOS-S (\timeform{0''.16} per a pixel) 
with that of the event images obtained by Suprime-Cam (\timeform{0''.20} per a pixel). 
We also convolve the $i$-band reference image with a Gaussian kernel 
to make the point spread function (PSF) size consistent with that of the event image. 

\begin{figure*}
 \begin{center}
  \includegraphics[width=15cm]{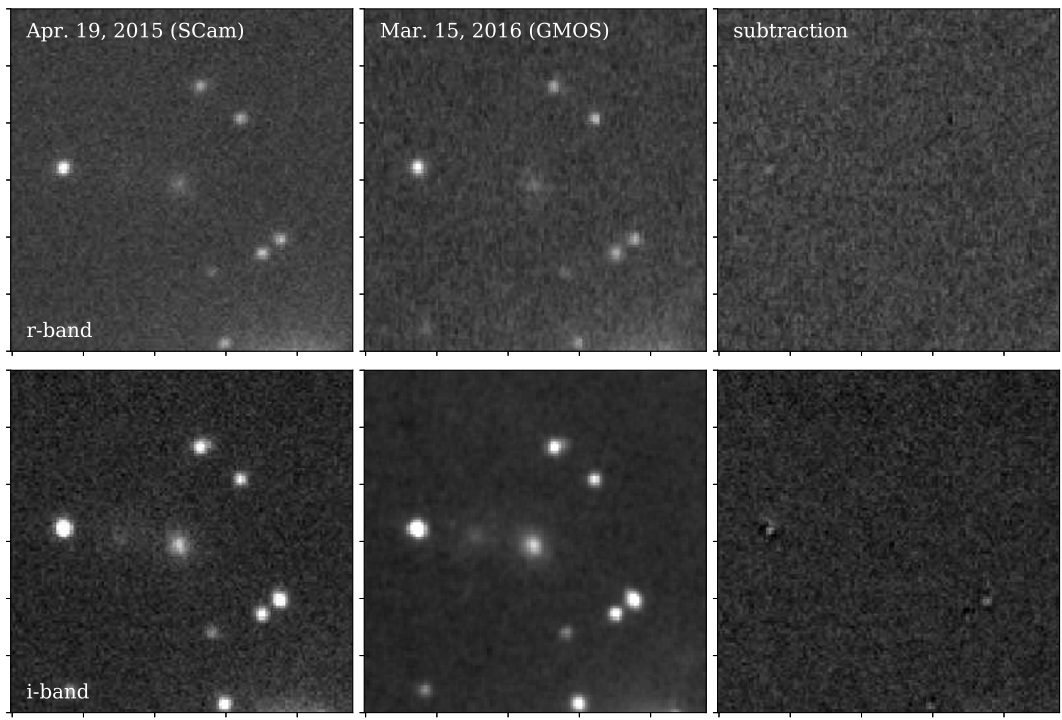}
 \end{center}
 \caption{
   {\it Left and middle panels}: same as Figure~\ref{fig:field} but zoomed into 
    a \timeform{24''} $\times$ \timeform{24''} region 
    centered on WISE~J0716--19. 
    The upper and lower panels are the images in $r$- and $i$-band, respectively. 
   {\it Right panels}: the subtraction of the reference images (the middle panels) 
    from the event images (the left panels). 
 }\label{fig:subtract}
\end{figure*}

The images of WISE~J0716--19 with the two filters at the two epochs 
and the subtracted images are shown in Figure~\ref{fig:subtract}. 
No residual source is visible at the position of WISE~J0716--19 in the subtracted images. 
To estimate the detection limits of the subtraction images, 
we randomly distribute a thousand circular apertures 
of \timeform{1".4} in diameter (twice the full width at half maximum of the PSF) 
on blank fields in the subtracted images, 
and investigate the distributions of the flux densities in those apertures. 
The $3\sigma$ scatter of the obtained distributions is $1.51\times10^{-30}$ 
and $1.65\times10^{-30}$ erg s$^{-1}$cm$^{-2}$Hz$^{-1}$ in $r$- and $i$-band, 
which we consider as the upper limits on a transient object 
that was occurring in WISE~J0716--19 at the time the event images were taken. 

To confirm the nonexistence of a variable source in WISE~J0716--19 quantitatively, 
we perform aperture photometry with circular apertures of \timeform{1".4} in diameter
at the position of WISE~J0716--19 on the subtracted images. 
The resulting flux densities are $2.64\times10^{-31}$ 
and $-1.06\times10^{-31}$ erg s$^{-1}$cm$^{-2}$Hz$^{-1}$ in $r$- and $i$-band,  
which is consistent with the limits derived above. 

\section{Discussion}

Taking account of the redshift $z=0.492$ of WISE~J0716--19 
and correcting for the large foreground extinction 
of $A_V = 3.7$ in the direction \citep{Schlafly:2011a}, 
the upper limits derived in the previous section correspond 
to absolute magnitudes of $> -19.4$ and $> -18.7$ 
at restframe wavelengths of 4200 and 5100 \AA , respectively. 
The absolute limiting magnitudes are fainter than 
peak magnitudes of type Ia supernovae (SNe) and broad-lined type Ic SNe, 
while they are brighter than most type II SNe even at the peak of their lightcurve 
(e.g., \cite{Barbary:2012a, Okumura:2014a, Whitesides:2017a, Dahlen:2012a}). 
However, the peak time of a SN lightcurve is typically $\sim 10$ days after the burst. 
Taking into account that the event images were taken 1 day after the occurrence of FRB~150418, 
association of a SN of any type with FRB~150418 is not ruled out 
even if FRB~150418 really occurred in WISE~J0716--19. 

Unlike a SN, an optical afterglow of a gamma-ray burst (GRB) 
usually reaches its peak luminosity within 1 day after the burst 
(for reviews of the observational properties 
of GRB optical afterglows, see \cite{Kann:2011a} and references therein). 
The absolute limiting magnitudes derived above 
is comparable to luminosities of the short GRB afterglows 1 day after the bursts 
(optical absolute magnitude $\sim -21$ to $-18$), 
and hence an afterglow could have been observed 
if a short GRB (or a long GRB whose afterglow is typically brighter) 
occurred in WISE~J0716--19 simultaneously with FRB~150418. 
It has also been pointed out that the energy 
of the outflowing material is comparable to that of a short GRB, 
if the ATCA object is a similar phenomenon as a GRB afterglow \citep{Zhang:2016a}. 
However, an afterglow would not be visible when the GRB event is off-axis. 
We also note that optical afterglows are not detected for many short GRBs, 
and the sample of short GRB afterglows with known luminosity 
may represent the bright end of the overall population. 
Thus, the occurrence of a GRB in WISE~J0716--19 is not ruled out. 

The radio emission of WISE~J0716--19 suggests that it 
hosts a radio faint AGN \citep{Williams:2016a, Vedantham:2016b, Bassa:2016a, Giroletti:2016a, Johnston:2017a}. 
However, the optical spectrum of WISE~J0716--19 shows 
no AGN signature \citep{Keane:2016a}, 
suggesting that the disk luminosity of any putative AGN is low. 
Our non-detection of any optical variability also supports this interpretation. 

The constraints on the optical variability of WISE~J0716--19 are weak 
largely due to the foreground extinction of $A_V = 3.7$. 
Optical follow up observations of FRBs at higher Galactic latitudes 
where extinction in the Milky Way is small are desired 
to search for an optical counterpart of a FRB. 

\begin{ack}
We thank the anonymous referee for his/her helpful comments. 
This work was supported by JSPS KAKENHI Grant Number JP17K14255.   
\end{ack}

\bibliographystyle{apj}
\bibliography{reference_list}

\end{document}